\documentstyle[11pt,leqno]{article}

 \newcommand{\beq}{\begin{equation}}                       
 \newcommand{\eeq}{\end{equation}}                         
 \newcounter{nt}[section]                                  
 \newcounter{nl}[section]                                  
 \date{}                                                   

\begin{document}

\noindent
{\Large{ \bf Diffusion and wave behaviour in linear Voigt model}}

\vspace{5mm}

{\sc \bf {Monica DE ANGELIS* - Pasquale RENNO*}}

 \vspace{3mm}

\noindent
{\footnotesize {\bf *Dipartimento di Matematica e Applicazioni. Facolt\`{a} di
             Ingegneria, via Claudio 21,

\noindent 80125, Napoli.
             E-mail modeange@unina.it}}


\vspace{7mm}

\footnoterule
\vspace{7mm}
{\small
\noindent
\parbox[b]{0.8in}{{\bf Abstract.}}
\parbox[t]{4.5in} {A  boundary value  problem $ {\cal P}_\varepsilon$ related to a third-
order parabolic equation
with a small parameter $\varepsilon  $ is analized. This equation models
the one-dimensional evolution of many dissipative media as
viscoelastic fluids or solids, viscous gases, superconducting
materials, incompressible and electrically conducting fluids.
Moreover,
 the third-order parabolic operator regularizes various non linear
second order wave equations. In this paper,
 the hyperbolic and parabolic behaviour of the solution of  $ {\cal P}_\varepsilon$ is estimated by means of
{\em slow time}
$ \tau =\varepsilon t$  and {\em fast time}  $ \theta
= t/ \varepsilon$. As consequence,
 a rigorous asymptotic approximation
 for the solution of  $ {\cal
P}_\varepsilon$ is established.

\vspace{3mm}
\noindent{\bf partial different equations / viscoelasticity/
superconductivity/ boundary layer}

\noindent
\vspace {3mm}

{\bf Diffusion et comportement ondullux dans le mod\'{e}l linear
de Voigt}
}}

\noindent
\vspace {3mm}

\noindent
\parbox[b]{0.8in}{{\bf R\'{e}sum\'{e}.}}
\parbox[t]{4.5in} {\small{{\em On analyse un probl\`{e}me  $ {\cal P}_\varepsilon$
 des valeurs au contour relativement \` {a} une equation parabolique du
 troisi\`{e}me ordre. Cette equation regle l'evolution unidimensionnel
 de beaucoup de materiels dissipatifs comme le fluides ou les solides
 visquelastiques, les gaz visqueux, les materiels superconductibles,
 les fluides incompressibles \'{e}lectriquement conductibles. De plus
 l'op\'{e}rateur parabolique du troisi\`{e}me ordre regularise divers
 equations non lineaires des ondes du deuxi\`{e}me ordre. On examine
 dans ce travail le comportment hyperbolique ou parabolique de la
 solution du  $ {\cal P}_\varepsilon$
moyennant le temps lent et le temps rapide. En cons\'{e}quence, on
pose une rigoureuse approximation asymptotique pour la solution du  $
{\cal P}_\varepsilon$.}}
\vspace{3mm}

\noindent{\bf \'{e}quations aux d\'{e}riv\'{e}es partielles /
viscollasticit\'{e} / supraconductivit\'{e} }}

\vspace{7mm}
\footnoterule
\vspace{7mm}
\noindent {\bf 1.}{ \bf Introduction}
\setcounter {section}{1}
\setcounter{equation}{0}

\hspace{5.1mm}

The parabolic equation
\beq                     \label{11}
 {\cal L}_ \varepsilon u \equiv \varepsilon \partial_{xxt} u
+ c^2 \partial_{xx}u  - \partial_{tt}u  = -f
\eeq

\noindent
describes a great deal of models of applied sciences and represents a typical example of hyperbolic equations perturbed by
viscous terms.

According to the meaning of $f $, examples of dissipative phenomena
related to (\ref{11}) are: motions of viscoelastic fluids
or solids \cite {jrs,M,r2}, heat conduction at low
temperature\cite{mps,fr},
sound propagation in viscous gases \cite {la}, propagation of
plane waves in perfect incompressible and electrically conducting
fluids \cite{na}. Moreover, when  $f = a u_t+ sin  u -\gamma$, the equation (\ref{11})
is the {\it
perturbed sine-Gordon equation}  which models the
Josephson tunnel effect in Superconductivity \cite {bp}. Further applications of
(\ref{11}) arise in the study of viscoelastic plates with memory, when
the relaxation function is given by an exponential function
 \cite{rf}.
At last, one remarks that even the
Navier- Stokes equations for a compressible gas with small viscosity,
in Lagrangian coordinates, can be reduced to (\ref{11}) with
$f=f(u_t,u_x,u_{xt}, u_{xx})$ \cite{mm}.

Also the meaningful analytical results
concerning the qualitative
analysis of (\ref{11}) are very
numerous and one can refer to an extensive bibliography
 (e.g. \cite{mm}-\cite{s}
).
In particular, the behaviour of solutions of (\ref{11}) when $\varepsilon \rightarrow
0$  has been analized in various applications of artificial viscosity
method to
non linear second order wave equation
 \cite
{kl},\cite{n}.
But, from a physical point of view, it would be interesting to
 estimate the
 time - intervals where the
hyperbolic or parabolic behaviour prevails, evaluating so the
influence of dissipative causes on the wave propagation.

These aspects are analyzed in this paper referring to the strip
problem
 ${ \cal P} _\varepsilon$ for  equation (\ref{11}) with a linear $f$.
The Green function $G$ related to  ${ \cal P} _\varepsilon$
 has already been
 determined in \cite{mda} by means of
 a rapidly decreasing Fourier series, and
its asymptotic behaviour for  $t \rightarrow
\infty$ has been obtained, too.

Now, in the hypotesis of $\varepsilon$ vanishing, appropriate estimates
of  $G$
 by the {\em  slow time} $\varepsilon t$ and the {\em fast time}
$t/\varepsilon$ will be established.
As consequence, the main result is a
rigorous approximation for the solution of the problem  ${ \cal P} _\varepsilon$
which holds for all
$ t < \varepsilon ^{-\eta} \
\ (\eta >0).$

\vspace{8mm}
\noindent
{\bf 2. Statement of the problem}
 \setcounter{section}{2}
 \setcounter{equation}{0}

 \hspace{5.1mm}

If $T$ is a positive constant and

\vspace{4mm}

$\ \ \ \ \ \ \ \ \ \ \ \  \ \ \ \ \ \ \   D =\{(x,t) : 0 \leq x \leq
l, \  \ 0 < t \leq T \}$,

\vspace{3mm}
\noindent
let $u(x,t) $ the regular solution of the boundary initial value
problem:

  \beq                                                     \label{21}
  \left \{
   \begin{array}{ll}
    & \partial_{xx}
    (\varepsilon u_t +
c^2 u) - \partial_{tt} u  = \ f(x,t),\ \  \
       (x,t)\in D,\vspace{2mm}\\
   & u (x,0)=f_0(x), \  \    u_t (x,0)=f_1(x),
   \ \ \  \ x\in [0,l],\vspace{2mm}  \\
    & u (0,t)=0, \  \ u (l,t)=0, \ \ \  \  0< t \leq T,
   \end{array}
  \right.
 \eeq

\noindent
where $f(x,t)$ is an arbitrary specified function.

Now, denote with $w(x,t)$ the solution of the reduced problem obtained
by (\ref{21}) with $\varepsilon=0$.
To establish a rigorous asymptotic approximation for $u(x,t) $ when $\varepsilon
\rightarrow 0$, we put:
\beq                                         \label{22}
u(x,t,\varepsilon)= e^{-\varepsilon t}  w(x,t) + r(x,t,\varepsilon)
\eeq

\noindent
where  the {\it error} $r(x,t,\varepsilon )$ must be estimated.

By means of standard computations one verifies that $r(x,t,\varepsilon)$
is the solution of the problem:

  \beq                                                     \label{23}
  \left \{
   \begin{array}{ll}
    & \partial_{xx}
    (\varepsilon r_t +
c^2 r) - \partial_{tt} r  = \ F(x,t,\varepsilon),\ \  \
       (x,t)\in D,\vspace{2mm}\\
   & r (x,0)=0, \  \    r_t (x,0)=0,
   \ \ \  \ x\in [0,l],\vspace{2mm}  \\
    & r (0,t)=0, \  \ r (l,t)=0, \ \ \  \  0< t \leq T,
   \end{array}
  \right.
 \eeq

\noindent
where the source term $F(x,t,\varepsilon)$ is:
\beq                                                \label{24}
F(x,t,\varepsilon)=f(x,t)(1-e^{-\varepsilon t})+ \ e^{-\varepsilon t}[-\varepsilon
\lambda_t+\varepsilon^2 (w+w_{xx})],
\eeq

\noindent
with  $\lambda= 2w+w_{xx}$.

The problem (\ref{23}) has already been solved in \cite{mda} and
the solution is given by:
\beq                                      \label{25}
r(x,t,\varepsilon)= -\int_{0}^{l} d\xi \\ \int_{0}^{t}
F(\xi,\tau,\varepsilon) \\\\\\ G(x,\xi,t-\tau) \ d\tau,
\eeq

\noindent
where $G(x,\xi,t)$ is:
\beq                                     \label{26}
G(x,\xi,t)=\frac{2}{l} \ \ \\ \ \sum_{n=1}^{\infty} \ \ H_n(t) \ \
\sin \gamma_n x  \  \sin \gamma_n \xi,
\eeq

\noindent
with
\vspace{3mm}
\beq                             \label{27}
H_n(t)= \ \ \frac{e^{-bn^{2}t}}{bn^2 \sqrt{1-(k/n)^2}} \
\sinh \{ bn^2t \sqrt{1-(k/n)^2}\},
\eeq

\noindent
and
\beq                 \label{28}
b= \frac{\pi^2}{2l^2}\varepsilon = q \varepsilon, \\\\ \ \ \ \ k=\frac{2cl}{\pi\varepsilon} \
\ \ \ \\\ \  \  \gamma_n =\frac{\pi}{l} n.
\eeq

\vspace{8mm}
\noindent
{\bf 3. Estimates of the Green function by fast and slow times}
 \setcounter{section}{3}
 \setcounter{equation}{0}

\hspace{5.1mm}

When  $\varepsilon \rightarrow 0 $, two characteristic times affect the
behaviour of $G$, i.e. $ \tau =\varepsilon t$ ({\em slow time}), and $ \theta
= t/ \varepsilon$ ({\em fast time}).
To point out the different contributions an
appropriate form of G we will considered.
For this, if
 $N= [\frac {2cl}{\pi \varepsilon}]$, the G-function can be given the
 form:
\beq                             \label{31}
G = \frac{2}{l} \ \ \\ \{  \sum_{n=1}^{N} \
+  \sum_{N +1}^{\infty} \\ \} H_n(t)
\sin(\gamma_nx)  \sin(\gamma_n\xi)   =  G_1 + G_2.
\eeq

\noindent
where, for $n<N,$ the functions $H_n$ are:
\beq                             \label{32}
H_n(t)= \ \ \frac{e^{-bn^{2}t}}{bn^2 \sqrt{(k/n)^2-1}} \
\sin \{ bn^2t \sqrt{(k/n)^2-1} \}.
\eeq

\noindent
If $\alpha $ is an arbitrary constant such that:
\beq                         \label{33}
1/2 < \alpha <1 , \ \ \ \ \  N_\alpha
=[ \frac{2cl}{\pi \varepsilon ^\alpha}],
\eeq

 \noindent
the term $G_1 $  can be written:
\beq                       \label{34}
G_1(x,\xi,t)= \frac{2}{l} \ \ \\ \ \{ \sum_{n=1}^{ N_\alpha} \  H_n(t) \
+  \sum_{ N \alpha+1}^{N} \\ H_n(t) \}
 \sin (\gamma_n x)  \ \sin (\gamma_n \xi) \ .
\eeq

\noindent
It is easy to prove that when $1\leq n \leq N_\alpha$ it results:
\vspace{2mm}
\beq                 \label{35}
\sqrt{(k/n)^2-1} \geq
\frac{\sqrt{1-\varepsilon^{2(1-\alpha)}}}{\varepsilon^{1-\alpha}}; \ \ \
 e^{-bn^{2}t} \leq  e^{-qt \varepsilon}.
\eeq

\vspace{2mm}
\noindent
Otherwise, if  $ N_\alpha+1 \leq n \leq N $, one has $N =
k-\beta$ with  $0<\beta<1$, and it is:
\vspace{2mm}
\beq                 \label{36}
\sqrt{(k/n)^2-1} \geq
\frac{ \sqrt{\pi\varepsilon \beta } \sqrt{4cl-\beta \pi \varepsilon}}{
 (2cl-\pi \varepsilon\beta)}; \ \ \
 e^{-bn^{2}t} \leq  e^{-2 c^2 t / \varepsilon^{2\alpha-1}}.
\eeq

\vspace{2mm}
\noindent
In particular, when $k$ is an integer one has
$\beta=1$ and the term $t
e^{-2c^2 t/\varepsilon}$ must be considered too.

For all  $ \varepsilon \in \ ]0,\varepsilon_0] \ (\varepsilon_0 <1)$, the
formulae
 (\ref{35}) - (\ref{36}) allow to obtain the following estimate for
 $G_1$:

\beq                                          \label{37}
|G_1(x,\xi,t)| \leq  A_0 \   \varepsilon^{-\alpha}
e^{-qt\varepsilon}   +  A_1 \ \varepsilon ^{-3/2} \
e^{-c^2t/\varepsilon^{2\alpha-1}},
\eeq

\noindent
 where the constants  $A_0 \ \ A_1 $ don't depend on $\varepsilon $ .
As $\varepsilon < \varepsilon^{1-2\alpha}$, the prevailing term in
(\ref{37}) is related to the slow time $\tau=\varepsilon t$. So, the
circular component $G_1$ is controlled by the slow time. On the
contrary, the hyperbolic component $G_2 $ is characterized only by the
fast time $\theta$.
 In fact, let:
\beq                        \label{38}
C=\frac{\pi (1-\beta)4cl }{2 cl+\pi
(1-\beta)}, \ \ \  C_1 =\frac{2 \zeta(2)}{qlC}
\eeq

\noindent
with $\beta \equiv 0$ if $k$ is an integer. Observing that
 $\forall n \geq  N+1$, one has (see, f.i. \cite{mda}):
\vspace{2mm}
\beq                     \label{39}
bn^2t (1\pm \sqrt{1-(k/
n)^{2}}) \geq c^2 t / \varepsilon,\ \ \sqrt{1-(k/n)^{2}} \geq
\varepsilon \  C,
\eeq

\vspace{2mm}
\noindent
and consequently it results:
\vspace{2mm}
\beq           \label{311}
|G_2(x,\xi,t)|  \leq C_1  \ \varepsilon^{-2} \
e^{-c^2t/\varepsilon}.
\eeq

\noindent
So, if   $M_0 = max \{A_1
, C_1  \}$ and  $ 1/2 < \alpha < 1  $, the following
theorem holds:

\vspace{3mm}
{\bf Theorem
3.1}
{\em For all} $\varepsilon \in (0, \varepsilon _0] (\varepsilon
_0 < 1 ) $ and $(x,t) \in D $, {\em the Green function} $G(x,\xi,t)  $
{\em verifies the following estimate}:
\vspace{2mm}
\beq                         \label{313}
|G(x,\xi,t)| \leq  A_0  \ \varepsilon^{-\alpha}
e^{-qt\varepsilon}   +  M_0 \ \varepsilon
^{-3/2}
e^{-c^2t/\varepsilon^{2\alpha-1}}.
\eeq

\noindent
{\em where the constants} $A_0, \ M_0$ {\em do not depend on} $\varepsilon$.
\hbox{} \hfill \rule {1.85mm}{2.82mm}

\vspace{8mm}
\noindent
{\bf 4. On the behaviour of the solution}
 \setcounter{section}{4}
 \setcounter{equation}{0}

 \hspace{5.1mm}

Now, the remainder term $r(x,t,\varepsilon)$ of (\ref{22}) can be
estimated. Referring to the function $f$ defined in
(\ref{24}), let

\vspace{3mm}

\beq                             \label{41}
 \ \ ||F||= max \{\sup_{D} \  |f(x,t)|,
\sup_{D} \  [ |\lambda_t| +\varepsilon |\lambda-u|] \  \}.
\eeq

\noindent
Then, one has the following theorem.

\vspace{5mm}
{\bf Theorem 4.1 -} {\em Let}  $F(x,t,\varepsilon)$
 $ \in C^1(D)$
{\em and let} $ F, F_x,F_t $  {\em bounded for all t.}
{\em  Then, the error term}  $r(x,t,\varepsilon)$ {\em verifies the estimate:  }

\beq                                         \label{42}
 |r(x,t,\varepsilon)| \ <  \ k  \  ||F|| \ (\varepsilon ^{ \eta}  \
 t)^2,
\eeq

\noindent
{\em where the constants} k {\em and} $\eta$ {\em do not depend on}
$\varepsilon $ {\em and}  $
\eta \in (0,1/2) $.

\vspace{3mm}
{\bf Proof}- First, by (\ref{24})-(\ref{25}) one deduces:
\beq                              \label{43}
| r(x,t,\varepsilon )| \leq l  \varepsilon \int _0^t e^{-\varepsilon \tau}
\{ |\lambda_t(x,\tau)|+ \varepsilon |\lambda -u|\}
|G(x,\xi,t-\tau)|  d\tau
\eeq
\hspace* {2cm} \[+ l  \int_0^t |f(x,\tau)|
|1-e^{-\varepsilon \tau} |G(x,\xi,t-\tau)| d\tau.\]

\noindent
Further, by means of the well- known
inequality \cite{m}:
\beq                           \label{44}
e^{-x} \leq [\gamma/(ex)]^\gamma    \ \ \   \forall \gamma>0, \forall x>0
\eeq

\noindent
by (\ref{311}) and (\ref{43}) one can deduce:
\beq                         \label{45}
|r| \leq \  ||F|| \ l \ t^2
\  \{ 3/2  \ A_0 \ \varepsilon^{1-\alpha}  +
\frac{2 M_0 }{(e c^2)^\gamma (1-\gamma)}
\frac{\varepsilon^{(2\alpha-1)\gamma}}{\sqrt{\varepsilon}} \}
\eeq

\noindent
So, if $\alpha$ and $\gamma$  are such that $3/4<\alpha<1$ and
$[2(2\alpha-1)]^{-1}< \gamma <1$, it suffices to put
\beq                      \label{48}
 2\eta=min \{ (2\alpha-1)\gamma-\frac{1}{2}, \ 1-\alpha \} ;
k= max \{ \frac{3}{2}A_0, \frac{2 M_0 }{(e c^2)^\gamma (1-\gamma)} \},
\eeq

\noindent
to deduce (\ref{42}).
\hbox{} \hfill \rule {1.85mm}{2.82mm}

As consequence of Theorem 4.1, finally we can observe that
:
\vspace {3mm}

 {\em When} $\varepsilon \rightarrow 0,$ {\em   the solution
of the  problem} (\ref{21}) {\em can be approximatated by means of the
following formula:
}
\vspace{3mm}
\beq                                         \label{49}
u(x,t,\varepsilon)= e^{-\varepsilon t}  w(x,t) + r(x,t,\varepsilon)
\eeq
\vspace{3mm}
\noindent
{\em where the error }$r(x,t,\varepsilon)$ {\em is bounded
for all } $t < (1/ \varepsilon) ^{\eta}.  $

\hspace{5.1mm}

\begin {center} {\bf References}
\end {center}

\end{document}